\documentclass[%
 reprint,
superscriptaddress,
 amsmath,amssymb,
 aps,
 prl,
floatfix
]{revtex4-1}
\usepackage{pdfpages}
\usepackage{pgffor}
\usepackage{graphicx}% Include figure files
\usepackage{dcolumn}% Align table columns on decimal point
\usepackage{bm}% bold math
\usepackage{hyperref}
\usepackage{xr}
\usepackage{physics}
\usepackage{chemformula}
\usepackage{siunitx}
\usepackage{color,soul}
\usepackage{float}

\newcommand{\bmk}{\bm{k}}

\renewcommand{\braket}[1]{\expval{#1}}

\makeatletter
\AtBeginDocument{\let\LS@rot\@undefined}
\makeatother

\begin{document}

\preprint{APS/123-QED}

    \title{Magnetic Bulk Photovoltaic Effect: Strong and Weak Field}
    % \thanks{A footnote to the article title}%

    \author{Zhenbang Dai}
	\affiliation{Department of Chemistry, University of Pennsylvania, Philadelphia, Pennsylvania 19104--6323, USA}
    \author{Andrew M. Rappe}%
	\affiliation{Department of Chemistry, University of Pennsylvania, Philadelphia, Pennsylvania 19104--6323, USA}

\date{\today}

\begin{abstract}
Shift current and ballistic current have been proposed to explain the bulk photovoltaic effect (BPVE), and there have been experiments designed  to separate the two mechanisms. 
These experiments are based on the assumption that under magnetic field, ballistic current can have a Hall effect while the shift current cannot, which is from some energy-scale arguments and has never been proven.
A recent work  [\textit{Phys. Rev. B}  \textbf{103}, 195203 (2021)] using quantum transport formalism achieves a conclusion that shift current indeed has a Hall current, seemingly contradicting the previous assumption and making the situation more confusing. 
%{\color{red} AMR: say what they conclude not just that it contradicts}
Moreover, the behavior of BPVE under strong magnetic field is still unexplored.
In this Letter, using a minimal 2D tight-binding model, we carry out a systematic numerical study of the BPVE under weak and strong magnetic field by treating the field in a non-perturbative way.
Our model clearly shows the appearance of the magnetically-induced ballistic current along the transverse direction, which agrees with the previous predictions, and interestingly a sizable longitudinal response of the shift current is also observed, a phenomenon that is not captured by any existing theories where the magnetic field is treated perturbatively.
More surprisingly, drastically different shift current is found in the strong-field regime, and the evolution from weak to strong field resembles a phase transition.
We hope that our work could resolve the debate over the behavior of BPVE under magnetic field, and the strong-field behavior of shift current is expected to inspire more studies on the relation between nonlinear optics and quantum geometry.

\begin{description}
\item[Keywords]
BPVE, shift current, ballistic current, magnetic field, Hall effect
\end{description}
\end{abstract}

\maketitle

\textit{Introduction.}~The bulk photovoltaic effect (BPVE) refers to the DC current generation from uniform light illumination in a homogeneous material lacking inversion symmetry, in contrast to the traditional photovoltaic effect which requires a heterojunction~\cite{Belinicher80p199, vonBaltz81p5590}. 
To explain this phenomenon, two major different but complementary mechanisms were proposed, namely, \textit{shift current} and \textit{ballistic current}~\cite{Belinicher88p29}. 
Ballistic current can be understood as a classical current which originates from the asymmetric momentum distribution of carriers after the light excitation, while shift current is a pure quantum effect and can be attributed to the wave packet evolution during the interband transition~\cite{vonBaltz81p5590, Young12p116601, Dai21p177403}.
Intuitively, shift current describes the coordinate shift in real-space during the transition~\cite{Young12p116601}.
Since these two theories were proposed, there have been experimental attempts  to separate and validate them based on the assumption that the ballistic current involves carrier movement and can thus give rise to a Hall current under magnetic field, whereas shift current is barely influenced by any realistic magnetic field~\cite{Burger19p5588, Burger20p081113}.

This assumption stems from an early work by Ivechenko {\it et\ al.}~\cite{Ivchenko84p55}~in which the response of shift current and ballistic current to magnetic field was discussed.
Due to the classical nature of ballistic current, a Lorentz force could be exerted on the charge carriers when applying a uniform and static magnetic field, diverting the ballistic current to the transverse direction and thus generating a Hall signal. 
Built on this idea and considering the electron-phonon coupling as the source of the asymmetric momentum distribution, it is straightforward to set up the Boltzmann transport equation to compute the Hall current of ballistic current.
For shift current, however, only a statement without further proof was made about its response to magnetic field, that is, when the cyclotron frequency of the magnetic field is much smaller than the difference between the photon energy and the band gap, the shift current would not be influenced. 
This statement was solely based on the energy scale of the magnetic field and the electronic system, so it could be misleading, as the magnetic field is known to have an impact on the phase of the wave function (Aharonov-Bohm effect, for example)~\cite{Thouless82p405}, which is directly related to the shift current that explicitly depends on the phase of the transition momentum matrix~\cite{Young12p116601}.
Thus, the previous claim about shift current responsiveness to magnetic field is elusive and remains controversial.  

Indeed, Hornung and von Baltz recently revisited this problem by using the quantum transport equation built upon non-equilibrium Green's functions formalism, where they seem to arrive at a conclusion that the shift current can also have a Hall-like response to magnetic field~\cite{Hornung21p195203}. 
Specifically, their way of incorporating the magnetic field is to add the induced phase to the Green's function expressed in the position representation, and then expand the phase factor in a Taylor series and retain the zeroth- and linear-order terms. 
Such treatment will generate a transverse current that is proportional to the non-magnetic shift current, which made those authors conclude that the shift current could have Hall signal. 
However, a close inspection of their result will reveal that this current is the same as the ``magnetically-induced photogalvanic effect" in Ivechenko et al.\ ~\cite{Ivchenko84p55}, which is attributed to the breaking of time-reversal symmetry that causes the asymmetry of the momentum distribution (characteristic of ballistic current). 
Thus, both works achieved the same insight that under magnetic field, a new current along the Hall direction will emerge that is of ballistic type.
We will call this new current \textit{magnetically-induced ballistic current} (or mag-BC) in this Letter to distinguish it from electron-phonon or electron-hole ballistic current~\cite{Dai21p177403, Dai21p235203}.
No further discussion and investigation has been previously made about how shift current itself changes under magnetic field. 

One common feature of the previous works is that they both consider the impact of magnetic field as a small perturbation. 
This could be problematic since even a small magnetic field could have a nontrivial influence on the electronic states; for example, a magnetic field gives rise to the famous fractal band structure known as the Hofstadter butterfly~\cite{Hofstadter76p2239}. 
Hence, one should be cautious when applying the conclusions made in the aforementioned works to experimental design, and it is crucial to examine the magnetic BPVE non-perturbatively so as to understand the applicability of such conclusions.
To this end, in this Letter we employ a non-perturbative approach to investigate the behavior of BPVE under weak and strong magnetic field.
Our study is carried out in a two-band tight-binding model, but the approach established here can be easily extended to \textit{ab initio} calculations for more realistic systems with the help of Wannierization~\cite{Marzari12p1419, Pizzi20p165902,Lado16p035023}.
The numerical results from our simulation clearly show the existence of transverse mag-BC under weak field, but more surprisingly, the shift current will also have an appreciable longitudinal %{\color{red}transverse and longitudinal} 
response to weak field. 
Moreover, the strong magnetic field will drastically change the shift current spectrum so that two electronic phases in terms of the BPVE can be identified. 
To the best of our knowledge, this is the first systematic numerical study on magnetic BPVE.

\begin{figure}
\centering
\includegraphics[width=85mm]{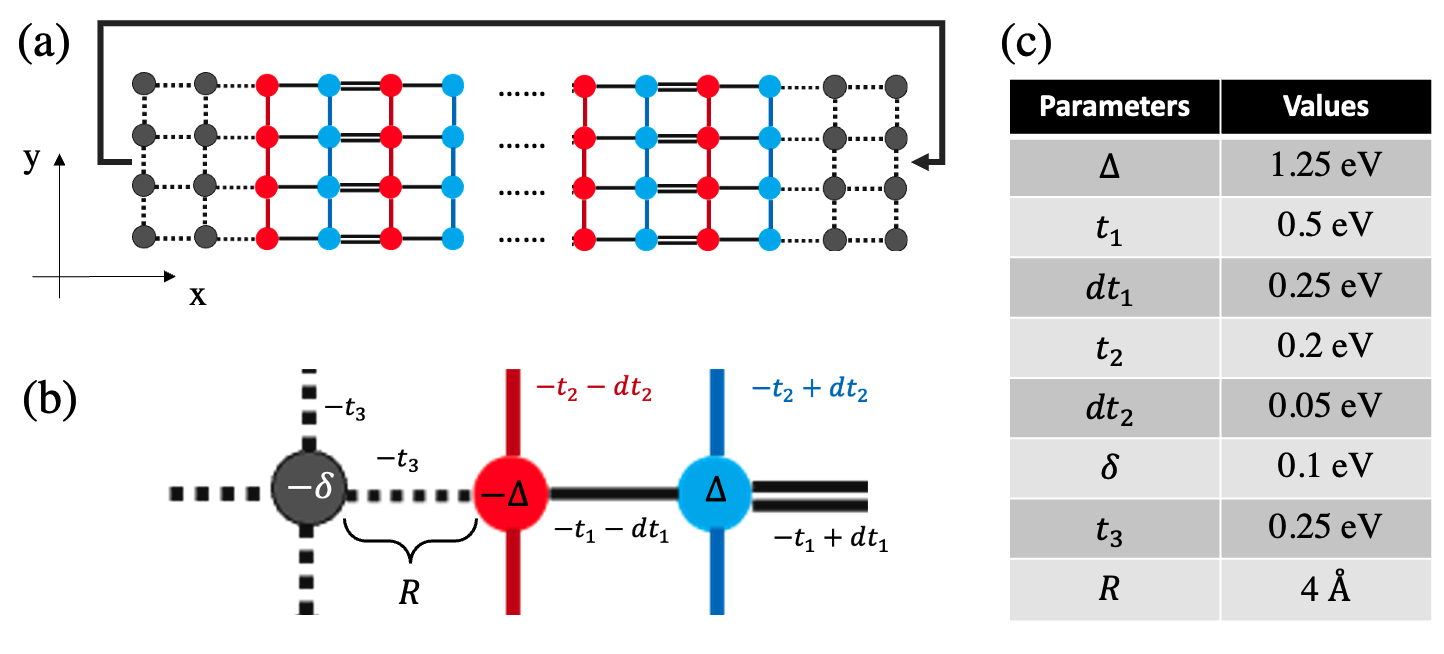}
\caption{\label{fig:model_system}
The model used in this work and the chosen parameters.
(a) The 2D tight-binding model used for calculating the magnetic BPVE. 
Blue and red atoms represent the optically active region while the gray atoms represent the leads.
The short-circuit condition is imposed by allowing  hopping between the two leads.
% Note that our model breaks inversion symmetry for staggered on-site energy and staggered hopping along the x-direction, which is required for BPVE.
(b) The parameters involved in the tight-binding model. 
(c) The values chosen for the parameters.
}
\end{figure}

\textit{Model.}~The simplest tight-binding model that breaks the inversion symmetry is the 1D Rice-Mele model (along $x$-direction, longitudinal) which has alternating onsite energies and hoppings~\cite{Rice82p1455, Ishizuka17p033015}. 
To study the Hall effect in a non-centrosymmetric system, we create a 2D tight-binding model by periodically repeating the long but finite Rice-Mele chain along the second dimension ($y$-direction, transverse) and adding a staggered hopping that connects them.
In order to model the short-circuit condition, two leads are attached to the ends of the Rice-Mele chain, and  hopping from one lead to the other is allowed~\cite{Kolpak06p054112}. 
All the bond lengths are chosen to be the same to simplify the model. 
The model can be seen in Fig~\ref{fig:model_system}a (the corresponding Hamiltonian can be found in SI), and the parameters and the chosen values for them in this work can be seen in Fig~\ref{fig:model_system}(b) and (c).
% The Hamiltonian for such model is:
% \begin{widetext}
% \begin{align}
%     \label{eqn:tb_model_1}
%     H=&H^{RM} + H^{L}, \nonumber \\
%     H^{RM}=&
%     \sum_{I, J} (-1)^{I} \Delta c_{I, J}^{\dagger}c_{I, J}
%     -\sum_{I, J}
%     \left(t_{1} + (-1)^{I} dt_1 \right)
%     c_{I+1, J}^{\dagger} c_{I, J}
%     -\sum_{I, J} 
%     (t_{2} + (-1)^I dt_2 )c_{I, J+1}^{\dagger} c_{I, J} 
%     + c.c.,
%     \\
%     \label{eqn:tb_model_2}
%     H^{L}=&-\sum_{K, J} \delta d_{K, J}^{\dagger}d_{K, J} 
%     - \sum_{K,J} t_3 d_{K+1, J}^{\dagger}d_{K, J} 
%     - \sum_{K, J} t_3 d_{K, J+1}^{\dagger}d_{K, J} 
%     - \sum_{\{I,K\},J} t_3 d_{K, J}^{\dagger}c_{I, J} + c.c..
% \end{align}
% \end{widetext}
% Eq.~(\ref{eqn:tb_model_1}) describes the central optically active region, where the first term is the onsite energy, the second term is the hopping along the $x$-direction, and the third term is the hopping along the  $y$-direction. 
% Eq.~(\ref{eqn:tb_model_2}) describes the leads, where each term has a similar meaning as in Eq.~(\ref{eqn:tb_model_1}), except the fourth term which describes the coupling between the leads and the central region.
% $I$ and $K$ are the site indices along the $x$-direction for central region and leads, respectively, and $J$ is the site index along $y$-direction.
% $\{I,K\}$ represents the sites at the interface.
In this work, there are 400 sites in the central region and 4 sites in the leads. 
Adding more atoms in the leads does not change the results (See SI).

To investigate the magnetic BPVE, we consider a static and uniform magnetic field perpendicular to the 2D system and focused on the optically active central region (extending the magnetic field to the leads will have no impact. See SI.). 
The influence of the magnetic field on the electronic structure can be incorporated via the so-called {Peierls substitution}~\cite{Peierls33p763, Hofstadter76p2239, Fradkin13field},
$t_{i j} \rightarrow t_{i j}  \exp \left(i \frac{e}{\hbar} \int_{\bm{R}_{\bm{i}}}^{\bm{R}_{\bm{j}}} \bm{A}_{B} \cdot d \bm{r}\right),$
where $\bm{A}_{B}$ is the vector potential of the magnetic field and $t_{ij}$ denotes the general hopping from lattice site $i$ to site $j$, and the integral follows the straight line from $i$ to $j$.
The Peierls substitution can be justified by doing the $\mathbf{p} \rightarrow \mathbf{p} - q\mathbf{A}_B$ replacement in the Hamiltonian and modifying the atomic orbitals by multiplying an extra phase $ \frac{e}{\hbar} \int_{\bm{R}}^{\bm{r}} \bm{A}_{B} \cdot d \bm{r}$ when deriving the tight-binding Hamiltonian.
If we use the Landau gauge $\bm{A}_B = (0, Bx, 0)$, then for our system the Peierls phase will be 
$\frac{eB}{\hbar}  \frac{\left(R_{j, y}-R_{i, y}\right)\left(R_{j, x}+R_{i, x}\right)}{2}.$
In this way, the influence of the magnetic field has been treated non-perturbatively. 
% This magnetic Hamiltonian is diagonalized, and the quantities such as velocity matrix and Berry connection can be computed.
%{\color{red}Does the shift current derivation prove that it is independent of Hamiltonian form? Why does B field affect only the wfs and energies but not the form of shift current tensor}
As the expression of shift current is invariant under the magnetic field (see~\cite{Gao21p042032} and SI), the shift current response can be calculated according to the formula\cite{Young12p116601}:
\begin{align}
    \label{eqn:shift_current}
    &j_{rsq}^{sh} =\frac{\pi e^3}{\hbar \omega^2} \sum_{n, l} \sum_{r,s} \int_{BZ} 
    \frac{d \bm{k}}{(2\pi)^3}
    \left(f_{l\bmk}-f_{n\bmk}\right) 
    v_{nl}^s(\bmk) v_{ln}^r(\bmk)
    \nonumber \\
    & \times\left[-\frac{\partial \phi_{n l}^{r}(\bm{k})}{\partial k_{q}}-\left[\chi_{l q}(\bm{k})-\chi_{n q}(\bm{k})\right]\right] 
    % \nonumber \\
    % & 
    % \times
    \delta\left(\varepsilon_{l\bmk}-{\varepsilon_{n\bmk}} \pm \hbar \omega\right).
\end{align}
Here, $\chi_{n q}(\bmk) \equiv \braket{u_{n\bmk} | i\partial_{k_q} u_{n\bmk}}$ is Berry connection, and $\phi_{n l}^{r}(\bm{k})$ is the phase of the velocity matrix $v_{nl}^r(\bmk) \equiv \braket{n\bmk | \hat{v}^r | l\bmk}$.
% {\color{red} isn't the superscript a q instead of r here?}
% {\color{red} Unclear} 
In addition to the shift current which comes from the off-diagonal elements of the density matrix, the diagonal part of the density matrix will also have a non-vanishing contribution, unlike the more well-studied situation where there exists time-reversal symmetry that causes the diagonal contribution to be exactly zero. 
% {\color{red} Are you saying that we apply the shift current formula itself gives you diagonal response? Or you take the trace of rho times velocity? Unclear.}
The newly emergent contribution from the diagonal part of the density matrix (asymmetric momentum distribution), as alluded above, is termed as mag-BC and can be calculated from the following formula~\cite{Zhang19p3783, Kraut79p1548}:
% {\color{red} Where does this come from? This paper should reveal the physics in eq 5 and this eq. Seems like you just grabbbed a formula. So we didn't derive this--already known?}
\begin{align}
    \label{eqn:inj}
    j^{inj}_{rsq} &=\frac{\pi e^3 \tau}{ {\hbar \omega}^2 }
    \Re \Bigg[\sum_{l,n} \sum_{r,s}
    \int_{BZ} \frac{d\bmk}{(2\pi)^3} (f_{l\bmk} - f_{n\bmk}) 
    \nonumber \\
    & \times
    v_{nl}^r(\bmk) v_{ln}^s(\bmk) v_{nn}^q(\bmk)
    \delta (\varepsilon_{n\bmk} - \varepsilon_{l\bmk} \pm \hbar \omega)
    \Bigg].
\end{align}

As we are interested in the BPVE from the central region, we manually exclude the contributions from the bands of the leads.
Moreover, we will restrict our discussion to the linear light-polarization along the $x$- (longitudinal) direction.
A test of our model is presented in SI, which shows excellent agreement of shift current and mag-BC obtained from a bulk system and from our model with leads, for both non-magnetic case and commensurate magnetic field (i.e., when the magnetic flux through the central region is equal to an integer number of flux quanta $h/e$).
% which shows that the non-magnetic shift current from this model agrees with that calculated from a bulk system.
% Moreover, in the case where the magnetic field is commensurate with the size of  the central region, i.e., the magnetic flux through the central region is equal to an integer number of flux quanta $h/e$, the Peierls phase will increase by 2$n\pi$ from one end to another, restoring periodic boundary conditions for the magnetic unit cell. 
% Thus, one can still  calculate the current  by adopting a periodic bulk system for specific magnetic field values. These results are in excellent agreement with the results from our model with leads
% % {\color{red} do you mean the model with leads?} 
% for both longitudinal shift current and transverse mag-BC.
% Note that we only show the $xxX$ shift current and $xxY$ mag-BC because other components can be proven to vanish (See SI).
After showing the validity of our model, we use this model to explore in more detail how BPVE behaves at weak and strong magnetic field.

% \begin{figure}
%     \centering
%     \includegraphics[width=85mm]{FIG2_v2.png}
%     \caption{Comparison of the various BPVE photocurrents from our model with electrodes and from the bulk system.
%     % {\color{red}x axis should say photon energy. "with leads"}
%     (a) and (b): The $xxX$ component of the shift current without magnetic field and with magnetic field.
%     (c) and (d): The $xxY$ component of the magnetically-induced ballistic current with magnetic field.
%     The magnetic field is set to be commensurate with the 400 sites in the optically active region.
%     The agreement between the leads model ((a) and (c)) and bulk system ((b) and (d)) is excellent, enabling us to investigate how BPVE reponse evolves as the magnetic field changes.
%     } 
%     \label{fig:benchmark}
% \end{figure}

%{\color{red} AMR to here. 6/11/22 6PM}

\begin{figure}
    \centering
    \includegraphics[width=85mm]{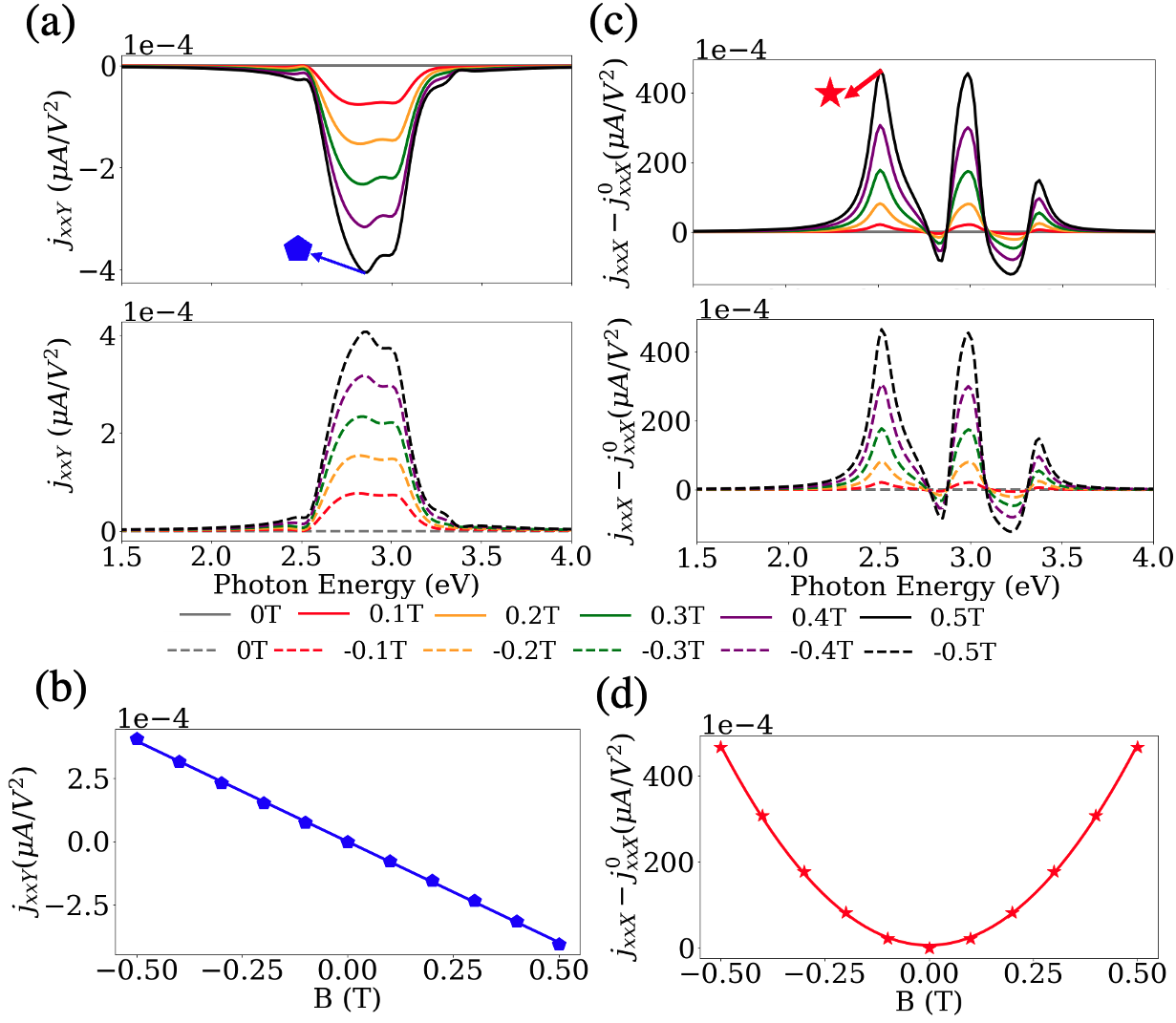}
    \caption{Magnetic BPVE in the weak-field regime. 
    (a) The response of the $xxY$ mag-BC with weak magnetic field. 
    (b) The scaling of mag-BC is linear in this regime for the selected peak (labeled by the blue pentagon in (a)).
    The existence and the scaling of mag-BC is consistent with the previous predictions~\cite{Hornung21p195203, Ivchenko84p55}, but its magnitude is very small and even smaller than the response of the shift current shown in (d).
    (c) The response of the $xxX$ shift current with weak magnetic field. 
    Different from previous predictions by~\cite{Hornung21p195203,Ivchenko84p55}, the shift current will respond to magnetic field quadratically, as can be seen from (d) for the selected peak (labeled by red star in (c)), even when the corresponding cyclotron frequency is much lower than the difference between photon energy and band gap.
    }
    \label{fig:result_weak_B}
\end{figure}

\textit{Weak magnetic field.}~Experimentally, a relatively weak magnetic field below 0.5~T is easily accessible, and it is in this regime where most of the controversy over the magnetic field influence on shift current occurs. 
So, it would be of interest to show the response of BPVE to a weak field.
%{\color{red} AMR to here. 6/12/22 10PM}
Shown in Fig.~\ref{fig:result_weak_B}a is the $xxY$ mag-BC (hereafter referred to as mag-BC) for magnetic field from -0.5~T to 0.5~T. 
When $B=0$~T, the mag-BC vanishes due to the existence of time-reversal symmetry, but once the magnetic field is turned on, mag-BC appears and scales linearly with the $B$ field strength, as can be seen in Fig.~\ref{fig:result_weak_B}b. 
% {\color{red} Fig 3 and all Figs have too small axis labels and headings and legends}
This is in agreement with the previous publications~\cite{Ivchenko84p55, Hornung21p195203} which also predicted a linear scaling of mag-BC $j^{mag-BC}_{\perp}=xBj_{\parallel}^{sh}$. (Though a close comparison between Fig.~S1 and Fig.~\ref{fig:result_weak_B}a reveals that the scaling coefficient $x$ is weakly frequency-dependent.)
Thus, in a sense mag-BC can be taken as a transverse response of shift current under the magnetic field, but it is not appropriate to interpret it as a classical Hall effect as will be discussed later.

To make connections with experiments~\cite{Burger19p5588, Burger20p081113} where a weak magnetic field was used to separate the shift current and ballistic current, it is worth noting that the mag-BC (Fig.~\ref{fig:result_weak_B}b) is five orders smaller
% {\color{red} tell readers what two things you are comparing}
than the non-magnetic shift current (Fig.~S1a and ~S1b), in contrast to the Hall effect of ballistic current 
% {\color{red} again, where should the reader look to get Hall effect of ballistic current?} 
which is only two orders smaller~\cite{Ivchenko84p55, Burger19p5588, Burger20p081113}.
In other words, the scaling coefficient $x$ should be much smaller than 1.
Therefore, one can safely neglect the contribution from mag-BC when analyzing the transverse current found in experiments with weak magnetic fields.

What is not captured by existing theories is that there is also a finite response of shift current to magnetic field~(Fig.~\ref{fig:result_weak_B}c and~\ref{fig:result_weak_B}d), which is even larger than mag-BC (Fig.~\ref{fig:result_weak_B}b. 
The response is symmetric with respect to positive and negative magnetic field, so it can only be described by even orders of response.
Such nonlinear behavior is reminiscent of the longitudinal current in classical Drude model for Hall effect, but unlike the Drude model, some peaks of the shift current will increase instead of decrease when the magnetic field is applied~\cite{Fradkin13field}.
This could be rationalized by the fact that the magnetic field will couple  states that are otherwise uncoupled (non-magnetic eigenstates having different crystal momenta, for example).
Thus, the longitudinal response of shift current to magnetic field is a distinct quantum effect and cannot be interpreted by classical models, and the failure to predict this response in~\cite{Ivchenko84p55, Hornung21p195203} again shows that the energy-scale argument is not sufficient to explain responses related to the quantum geometry.
% , and it might not be appropriate to treat the magnetic field as a small quantity and thus a perturbation in this case.
Moreover, the increasing magnitude of the longitudinal shift current also makes the interpretation of mag-BC as the classical Hall effect of shift current\cite{Hornung21p195203}, i.e. the diversion of shift current to the transverse direction, questionable since the diversion would make the shift current smaller.
As a result, it is more appropriate to interpret mag-BC as a new type of ballistic current originating from the asymmetric scattering due to the magnetic field.

\begin{figure}
    \centering
    \includegraphics[width=0.40\textwidth]{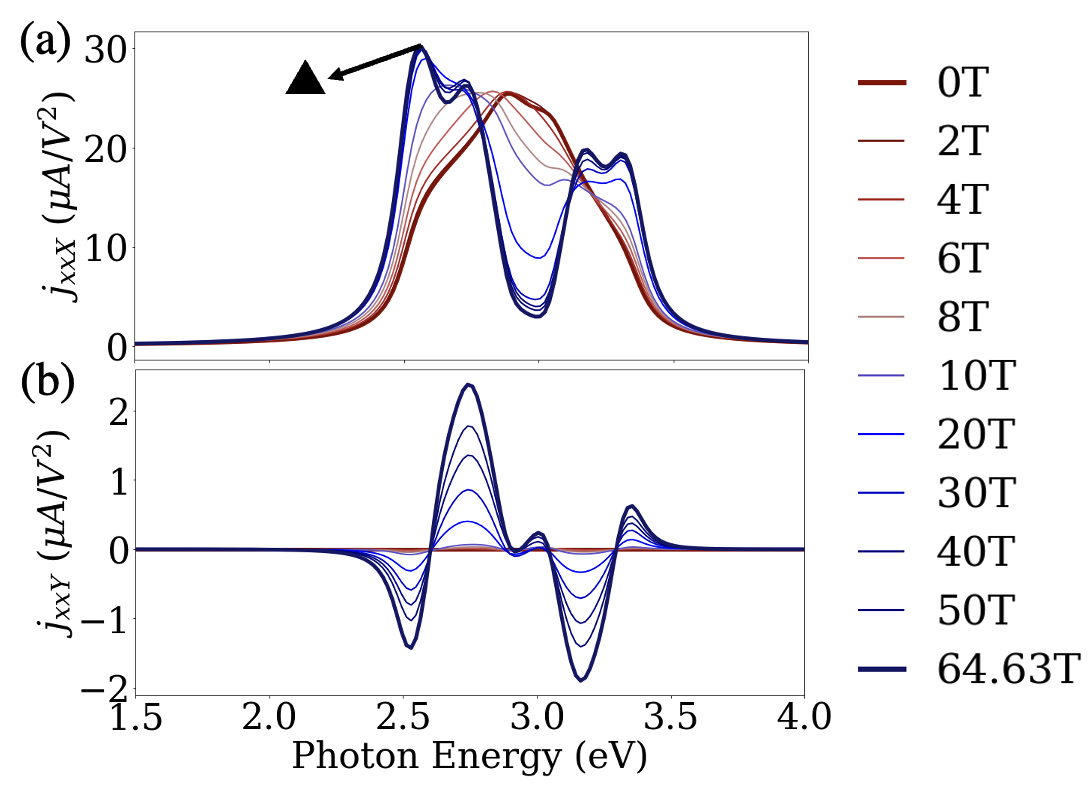}
    \caption{The transition of the magnetic BPVE from weak field to strong field.
    (a) The evolution of $xxX$ shift current. 
    % The shift current seems to experience a phase transition.
    Below 8~T, the profile is relatively constant but continues to grow quadratically. Above 20~T, the profile is also relative constant, though the lineshape is significantly different.
    Between 8~T and 20~T there is an intermediate regime where the low-field profile will transition into the strong-field profile.
    The evolution of the peak labeled by the black triangle is plotted in Fig.~\ref{fig:band_structure}a.
    (b) The evolution of $xxY$ mag-BC. Unlike the shift current, the mag-BC does not stabilize into two phases but continues to grow as the magnetic field becomes larger (providing stronger time-reversal symmetry breaking).
    % Some peaks (2.55~eV, 3.2~eV, 3.4~eV) always scale linearly with the magnetic field, whereas the peaks between 2.55~eV and 3.2~eV will shift their positions and change the trend as a result of the significant modification of the electronic structure induced by the strong magnetic field.
    }
    \label{fig:result_strong_B}
\end{figure}

\begin{figure}
    \centering
    \includegraphics[width=85mm]{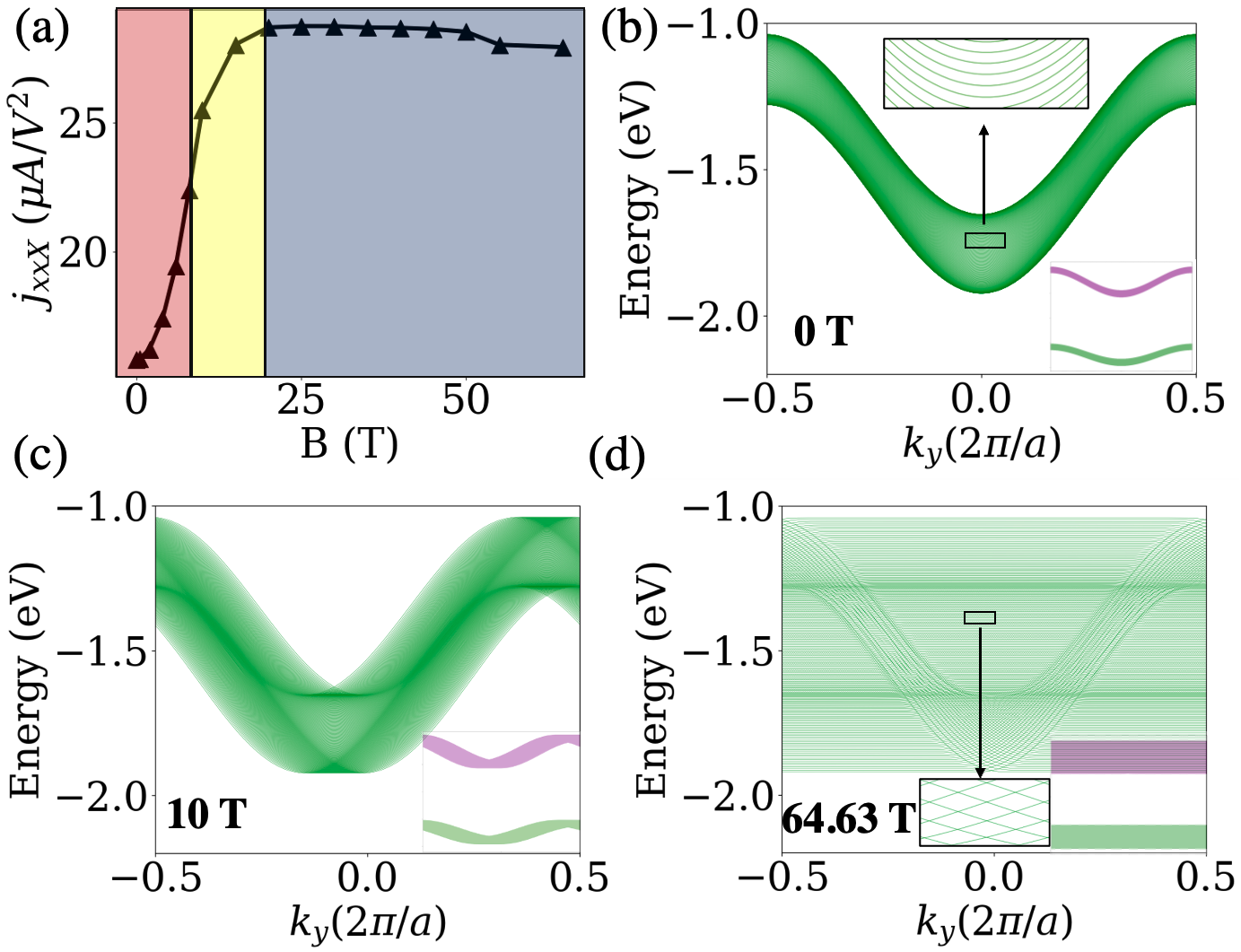}
    \caption{Evolution of the band structure with magnetic field.
    (a) The scaling of the shift current response represented by the black triangle in Fig.~\ref{fig:result_strong_B}a with magnetic field. 
    (b) The valence bands for the non-magnetic system. 
    % All the bands are well-separated, as can be seen in the black box. The bands will remain largely separate in the weak-field regime and the shift current will stay relatively constant.
    (c) The valence bands with 10~T magnetic field. 
    % Entering the intermediate regime, a significant portion of the bands have crossed each other, and accordingly, the shift current will begin to deviate from the non-magnetic profile.
    (d) The valence bands under the commensurate magnetic field (64.63~T). 
    % In the strong-field regime, band crossing happens in  most  of the Brillouin zone, making each electronic state have rather different character than the non-magnetic electronic states.
    % This large-scale change of band character is responsible for the different shift current profile at the strong field.
    For (b)-(d), the insets show both the valence band (green) and conduction band (purple) manifold, and the bands corresponding to leads have been removed. 
    }
    \label{fig:band_structure}
\end{figure}

\textit{Strong magnetic field.}~Comparing the non-magnetic shift current with the same quantity under the commensurate magnetic field (64.63~T) in Fig.~S1a, raises the question of how the spectrum evolves with increasing $B$ field, since it appears that the weak-field response shown in Fig.~\ref{fig:result_weak_B}c will not lead simply to the strong-field response. 
Thus, we further increase the magnetic field gradually and observe the change of the shift current lineshape.
The numerical results indicate that the shift current experiences a phase-transition-like change as shown in Figs.~\ref{fig:result_strong_B}a and~\ref{fig:band_structure}a.
Below 8~T, the lineshape of shift current is relatively constant despite the sizable quadratic response shown in Fig.~\ref{fig:result_weak_B}c, while between 8~T and 20~T, the shift current experiences rapid changes so that its lineshape begins to show the essential features of the strong-field response. 
Above 20~T, however, the lineshape again remains relative constant and exhibits a weak dependence on the magnetic field. 

To our knowledge, such peculiarity in the magnetic-field-induced change in BPVE has not been previously reported. Since  magnetic field affects the wave-vector-dependent wave function phases, these findings could be viewed as an example of how nonlinear optics can be used to probe the quantum geometry~\cite{Morimoto16pe1501524}.
This can be corroborated by the behavior of the $xx$ component of the absorption coefficient, which does not depend on the phase of the wave functions and thus shows negligible change as magnetic field varies (See SI). 
More concretely, it is the rapid change of band character that is responsible for the evolution of shift current lineshape~\cite{Tan16p237402}.
In the weak-field regime, all bands are well-separated from each other (Fig.~\ref{fig:band_structure}b), so that the bands have distinct band character; that is, each band can be largely ascribed to the delocalized states along $x$- and $y$- directions as in Fig.~\ref{fig:model_system}. 
As the magnetic field increases and exceeds a certain threshold, the well-separated bands start to mix and cross at some regions of the Brillouin zone (Fig.~\ref{fig:band_structure}c). 
Such band crossing is accompanied by a rapid change of the band character, which manifests as the fast change of the shift current lineshape, because the shift vector in Eq.~(\ref{eqn:shift_current}) depends sensitively on band characters~\cite{Tan16p237402}.
Entering into the strong-field regime, the band structure has a drastically different profile and band crossings happen almost everywhere in the Brillouin zone (Fig.~\ref{fig:band_structure}c). 
The band characters are thus totally changed compared with the non-magnetic bands, and therefore the shift current exhibits a completely different lineshape from the non-magnetic counterpart.

Contrary to the shift current, mag-BC is related to the asymmetry of the carrier generation rate, which is a result of the breaking of time-reversal symmetry caused by magnetic field.
So the mag-BC is more well-behaved in the sense that it will appear and become larger for stronger magnetic field (Fig.~\ref{fig:result_strong_B}b), as a stronger field strength would break the time-reversal-symmetry to a larger extent.
On the other hand, mag-BC can also feel the evolution of the band structure (Fig.~\ref{fig:band_structure}b and Fig.~S5) so that the peaks between 2.55--3.2~eV will shift their position somewhat, and so the trends of the  magnitudes at any one photon energy are not monotonic.
Nonetheless, the mag-BC will not stabilize into a constant lineshape as the shift current does and is always responsive to magnetic field.

\textit{Summary.}
In this Letter, we report a systematic
% the {\color{red} first-ever...I think we cannot claim novelty in papers} 
numerical study of the bulk photovoltaic effect under both weak and strong magnetic field. 
Our results clearly demonstrate the existence of transverse magnetically-induced ballistic current as predicted by several previous works, and a sizable longitudinal response from shift current to magnetic field is also revealed. 
Moreover, an interesting and surprising behavior for shift current when going from weak-field to strong-field is discovered, showing a phase-transition-like change in the lineshape.
We expect that this finding will inspire more studies on the relations between quantum geometry and nonlinear optics.

As a final note, we want to point out that the approach developed here, which is based on a minimal tight-binding model, can be extended to \textit{ab initio} calculations to investigate magnetic properties for real materials. 
For example, the Bloch states from mean-field calculations (Density functional theory, Hartree-Fock, etc.) can be Wannierized to give a fully first-principles tight binding model. 
Then, the Peierls phase could be added to hopping parameters in a similar fashion. 
The resulting Hamiltonian treats the magnetic field non-perturbatively, and it can thus help understand and predict phenomena whose important features could be hardly captured by perturbative approaches. 

% {\color{red} I will ask again. I think you are showing that the B field changes the wfs and the energies. But it seems you are assuming or proving that the response functions (shift current and injection current) formulas do not change. Is this assumed or proved? Is this right? Discuss with me.

% Explain in fig captions: blue pentagon, red star, blue triangle in Figures 3 and 4

% Does our work let us comment on whether the mag-BC should be considered to be proportional to shift current? I know you say x is frequency dependent, but does it basically not have a meaningful dependence on J_shift?

% Does the magnetically induced shift current always only have longitudinal component? It seems in this paper you have mag-BC and mag-shift in different directions. Is this always true?

% Can you fix the citations such as LYANDAGELLER? Dai Rappe no longer Arxiv? etc.}
% \section{Acknowledgement}

{\em Acknowledgments.} 
We acknowledge valuable discussions with Aaron M. Schankler and Dr. Lingyuan Gao. 
The work is supported by the U.S. Department of Energy, Office of Science, Basic Energy Sciences, under Award \# DE-FG02-07ER46431.

% \bibliography{BC}% Produces the bibliography via BibTeX.

\bibliography{rappecites, rappecite_temp}

%merlin.mbs apsrev4-1.bst 2010-07-25 4.21a (PWD, AO, DPC) hacked
%Control: key (0)
%Control: author (8) initials jnrlst
%Control: editor formatted (1) identically to author
%Control: production of article title (-1) disabled
%Control: page (0) single
%Control: year (1) truncated
%Control: production of eprint (0) enabled
\begin{thebibliography}{25}%
\makeatletter
\providecommand \@ifxundefined [1]{%
 \@ifx{#1\undefined}
}%
\providecommand \@ifnum [1]{%
 \ifnum #1\expandafter \@firstoftwo
 \else \expandafter \@secondoftwo
 \fi
}%
\providecommand \@ifx [1]{%
 \ifx #1\expandafter \@firstoftwo
 \else \expandafter \@secondoftwo
 \fi
}%
\providecommand \natexlab [1]{#1}%
\providecommand \enquote  [1]{``#1''}%
\providecommand \bibnamefont  [1]{#1}%
\providecommand \bibfnamefont [1]{#1}%
\providecommand \citenamefont [1]{#1}%
\providecommand \href@noop [0]{\@secondoftwo}%
\providecommand \href [0]{\begingroup \@sanitize@url \@href}%
\providecommand \@href[1]{\@@startlink{#1}\@@href}%
\providecommand \@@href[1]{\endgroup#1\@@endlink}%
\providecommand \@sanitize@url [0]{\catcode `\\12\catcode `\$12\catcode
  `\&12\catcode `\#12\catcode `\^12\catcode `\_12\catcode `\%12\relax}%
\providecommand \@@startlink[1]{}%
\providecommand \@@endlink[0]{}%
\providecommand \url  [0]{\begingroup\@sanitize@url \@url }%
\providecommand \@url [1]{\endgroup\@href {#1}{\urlprefix }}%
\providecommand \urlprefix  [0]{URL }%
\providecommand \Eprint [0]{\href }%
\providecommand \doibase [0]{http://dx.doi.org/}%
\providecommand \selectlanguage [0]{\@gobble}%
\providecommand \bibinfo  [0]{\@secondoftwo}%
\providecommand \bibfield  [0]{\@secondoftwo}%
\providecommand \translation [1]{[#1]}%
\providecommand \BibitemOpen [0]{}%
\providecommand \bibitemStop [0]{}%
\providecommand \bibitemNoStop [0]{.\EOS\space}%
\providecommand \EOS [0]{\spacefactor3000\relax}%
\providecommand \BibitemShut  [1]{\csname bibitem#1\endcsname}%
\let\auto@bib@innerbib\@empty
%</preamble>
\bibitem [{\citenamefont {Belinicher}\ and\ \citenamefont
  {Sturman}(1980)}]{Belinicher80p199}%
  \BibitemOpen
  \bibfield  {author} {\bibinfo {author} {\bibfnamefont {V.~I.}\ \bibnamefont
  {Belinicher}}\ and\ \bibinfo {author} {\bibfnamefont {B.~I.}\ \bibnamefont
  {Sturman}},\ }\href@noop {} {\bibfield  {journal} {\bibinfo  {journal} {Sov.
  Phys. USP.}\ }\textbf {\bibinfo {volume} {23}},\ \bibinfo {pages} {199}
  (\bibinfo {year} {1980})}\BibitemShut {NoStop}%
\bibitem [{\citenamefont {von{ }Baltz}\ and\ \citenamefont
  {Kraut}(1981)}]{vonBaltz81p5590}%
  \BibitemOpen
  \bibfield  {author} {\bibinfo {author} {\bibfnamefont {R.}~\bibnamefont {von{
  }Baltz}}\ and\ \bibinfo {author} {\bibfnamefont {W.}~\bibnamefont {Kraut}},\
  }\href@noop {} {\bibfield  {journal} {\bibinfo  {journal} {Phys.\ Rev.\ B}\
  }\textbf {\bibinfo {volume} {23}},\ \bibinfo {pages} {5590} (\bibinfo {year}
  {1981})}\BibitemShut {NoStop}%
\bibitem [{\citenamefont {Belinicher}(1988)}]{Belinicher88p29}%
  \BibitemOpen
  \bibfield  {author} {\bibinfo {author} {\bibfnamefont {V.~I.}\ \bibnamefont
  {Belinicher}},\ }\href@noop {} {\bibfield  {journal} {\bibinfo  {journal}
  {Ferroelectrics}\ }\textbf {\bibinfo {volume} {83}},\ \bibinfo {pages} {29}
  (\bibinfo {year} {1988})}\BibitemShut {NoStop}%
\bibitem [{\citenamefont {Young}\ and\ \citenamefont
  {Rappe}(2012)}]{Young12p116601}%
  \BibitemOpen
  \bibfield  {author} {\bibinfo {author} {\bibfnamefont {S.~M.}\ \bibnamefont
  {Young}}\ and\ \bibinfo {author} {\bibfnamefont {A.~M.}\ \bibnamefont
  {Rappe}},\ }\href@noop {} {\bibfield  {journal} {\bibinfo  {journal} {Phys.
  Rev. Lett.}\ }\textbf {\bibinfo {volume} {109}},\ \bibinfo {pages} {116601}
  (\bibinfo {year} {2012})}\BibitemShut {NoStop}%
\bibitem [{\citenamefont {Dai}\ \emph {et~al.}(2021)\citenamefont {Dai},
  \citenamefont {Schankler}, \citenamefont {Gao}, \citenamefont {Tan},\ and\
  \citenamefont {Rappe}}]{Dai21p177403}%
  \BibitemOpen
  \bibfield  {author} {\bibinfo {author} {\bibfnamefont {Z.}~\bibnamefont
  {Dai}}, \bibinfo {author} {\bibfnamefont {A.~M.}\ \bibnamefont {Schankler}},
  \bibinfo {author} {\bibfnamefont {L.}~\bibnamefont {Gao}}, \bibinfo {author}
  {\bibfnamefont {L.~Z.}\ \bibnamefont {Tan}}, \ and\ \bibinfo {author}
  {\bibfnamefont {A.~M.}\ \bibnamefont {Rappe}},\ }\href@noop {} {\bibfield
  {journal} {\bibinfo  {journal} {Physical Review Letters}\ }\textbf {\bibinfo
  {volume} {126}},\ \bibinfo {pages} {177403} (\bibinfo {year}
  {2021})}\BibitemShut {NoStop}%
\bibitem [{\citenamefont {Burger}\ \emph {et~al.}(2019)\citenamefont {Burger},
  \citenamefont {Agarwal}, \citenamefont {Aprelev}, \citenamefont {Schruba},
  \citenamefont {Gutierrez-Perez}, \citenamefont {Fridkin},\ and\ \citenamefont
  {Spanier}}]{Burger19p5588}%
  \BibitemOpen
  \bibfield  {author} {\bibinfo {author} {\bibfnamefont {A.~M.}\ \bibnamefont
  {Burger}}, \bibinfo {author} {\bibfnamefont {R.}~\bibnamefont {Agarwal}},
  \bibinfo {author} {\bibfnamefont {A.}~\bibnamefont {Aprelev}}, \bibinfo
  {author} {\bibfnamefont {E.}~\bibnamefont {Schruba}}, \bibinfo {author}
  {\bibfnamefont {A.}~\bibnamefont {Gutierrez-Perez}}, \bibinfo {author}
  {\bibfnamefont {V.~M.}\ \bibnamefont {Fridkin}}, \ and\ \bibinfo {author}
  {\bibfnamefont {J.~E.}\ \bibnamefont {Spanier}},\ }\href@noop {} {\bibfield
  {journal} {\bibinfo  {journal} {Science advances}\ }\textbf {\bibinfo
  {volume} {5}},\ \bibinfo {pages} {eaau5588} (\bibinfo {year}
  {2019})}\BibitemShut {NoStop}%
\bibitem [{\citenamefont {Burger}\ \emph {et~al.}(2020)\citenamefont {Burger},
  \citenamefont {Gao}, \citenamefont {Agarwal}, \citenamefont {Aprelev},
  \citenamefont {Spanier}, \citenamefont {Rappe},\ and\ \citenamefont
  {Fridkin}}]{Burger20p081113}%
  \BibitemOpen
  \bibfield  {author} {\bibinfo {author} {\bibfnamefont {A.~M.}\ \bibnamefont
  {Burger}}, \bibinfo {author} {\bibfnamefont {L.}~\bibnamefont {Gao}},
  \bibinfo {author} {\bibfnamefont {R.}~\bibnamefont {Agarwal}}, \bibinfo
  {author} {\bibfnamefont {A.}~\bibnamefont {Aprelev}}, \bibinfo {author}
  {\bibfnamefont {J.~E.}\ \bibnamefont {Spanier}}, \bibinfo {author}
  {\bibfnamefont {A.~M.}\ \bibnamefont {Rappe}}, \ and\ \bibinfo {author}
  {\bibfnamefont {V.~M.}\ \bibnamefont {Fridkin}},\ }\href@noop {} {\bibfield
  {journal} {\bibinfo  {journal} {Physical Review B}\ }\textbf {\bibinfo
  {volume} {102}},\ \bibinfo {pages} {081113} (\bibinfo {year}
  {2020})}\BibitemShut {NoStop}%
\bibitem [{\citenamefont {Ivchenko}\ \emph {et~al.}(1984)\citenamefont
  {Ivchenko}, \citenamefont {Lyandageller}, \citenamefont {Pikus},\ and\
  \citenamefont {Rasulov}}]{Ivchenko84p55}%
  \BibitemOpen
  \bibfield  {author} {\bibinfo {author} {\bibfnamefont {E.~L.}\ \bibnamefont
  {Ivchenko}}, \bibinfo {author} {\bibfnamefont {Y.~B.}\ \bibnamefont
  {Lyandageller}}, \bibinfo {author} {\bibfnamefont {G.~E.}\ \bibnamefont
  {Pikus}}, \ and\ \bibinfo {author} {\bibfnamefont {R.~Y.}\ \bibnamefont
  {Rasulov}},\ }\href@noop {} {\bibfield  {journal} {\bibinfo  {journal}
  {Soviet Physics Semiconductors-USSR}\ }\textbf {\bibinfo {volume} {18}},\
  \bibinfo {pages} {55} (\bibinfo {year} {1984})}\BibitemShut {NoStop}%
\bibitem [{\citenamefont {Thouless}\ \emph {et~al.}(1982)\citenamefont
  {Thouless}, \citenamefont {Kohmoto}, \citenamefont {Nightingale},\ and\
  \citenamefont {den Nijs}}]{Thouless82p405}%
  \BibitemOpen
  \bibfield  {author} {\bibinfo {author} {\bibfnamefont {D.~J.}\ \bibnamefont
  {Thouless}}, \bibinfo {author} {\bibfnamefont {M.}~\bibnamefont {Kohmoto}},
  \bibinfo {author} {\bibfnamefont {M.~P.}\ \bibnamefont {Nightingale}}, \ and\
  \bibinfo {author} {\bibfnamefont {M.}~\bibnamefont {den Nijs}},\ }\href@noop
  {} {\bibfield  {journal} {\bibinfo  {journal} {Physical Review Letters}\
  }\textbf {\bibinfo {volume} {49}},\ \bibinfo {pages} {405} (\bibinfo {year}
  {1982})}\BibitemShut {NoStop}%
\bibitem [{\citenamefont {Hornung}\ and\ \citenamefont {von
  Baltz}(2021)}]{Hornung21p195203}%
  \BibitemOpen
  \bibfield  {author} {\bibinfo {author} {\bibfnamefont {D.}~\bibnamefont
  {Hornung}}\ and\ \bibinfo {author} {\bibfnamefont {R.}~\bibnamefont {von
  Baltz}},\ }\href@noop {} {\bibfield  {journal} {\bibinfo  {journal} {Physical
  Review B}\ }\textbf {\bibinfo {volume} {103}},\ \bibinfo {pages} {195203}
  (\bibinfo {year} {2021})}\BibitemShut {NoStop}%
\bibitem [{\citenamefont {Dai}\ and\ \citenamefont
  {Rappe}(2021)}]{Dai21p235203}%
  \BibitemOpen
  \bibfield  {author} {\bibinfo {author} {\bibfnamefont {Z.}~\bibnamefont
  {Dai}}\ and\ \bibinfo {author} {\bibfnamefont {A.~M.}\ \bibnamefont
  {Rappe}},\ }\href@noop {} {\bibfield  {journal} {\bibinfo  {journal}
  {Physical Review B}\ }\textbf {\bibinfo {volume} {104}},\ \bibinfo {pages}
  {235203} (\bibinfo {year} {2021})}\BibitemShut {NoStop}%
\bibitem [{\citenamefont {Hofstadter}(1976)}]{Hofstadter76p2239}%
  \BibitemOpen
  \bibfield  {author} {\bibinfo {author} {\bibfnamefont {D.~R.}\ \bibnamefont
  {Hofstadter}},\ }\href@noop {} {\bibfield  {journal} {\bibinfo  {journal}
  {Physical review B}\ }\textbf {\bibinfo {volume} {14}},\ \bibinfo {pages}
  {2239} (\bibinfo {year} {1976})}\BibitemShut {NoStop}%
\bibitem [{\citenamefont {Marzari}\ \emph {et~al.}(2012)\citenamefont
  {Marzari}, \citenamefont {Mostofi}, \citenamefont {Yates}, \citenamefont
  {Souza},\ and\ \citenamefont {Vanderbilt}}]{Marzari12p1419}%
  \BibitemOpen
  \bibfield  {author} {\bibinfo {author} {\bibfnamefont {N.}~\bibnamefont
  {Marzari}}, \bibinfo {author} {\bibfnamefont {A.~A.}\ \bibnamefont
  {Mostofi}}, \bibinfo {author} {\bibfnamefont {J.~R.}\ \bibnamefont {Yates}},
  \bibinfo {author} {\bibfnamefont {I.}~\bibnamefont {Souza}}, \ and\ \bibinfo
  {author} {\bibfnamefont {D.}~\bibnamefont {Vanderbilt}},\ }\href {\doibase
  10.1103/RevModPhys.84.1419} {\bibfield  {journal} {\bibinfo  {journal}
  {Reviews of Modern Physics}\ }\textbf {\bibinfo {volume} {84}},\ \bibinfo
  {pages} {1419} (\bibinfo {year} {2012})}\BibitemShut {NoStop}%
\bibitem [{\citenamefont {Pizzi}\ \emph {et~al.}(2020)\citenamefont {Pizzi},
  \citenamefont {Vitale}, \citenamefont {Arita}, \citenamefont {Blügel},
  \citenamefont {Freimuth}, \citenamefont {G{\'{e}}ranton}, \citenamefont
  {Gibertini}, \citenamefont {Gresch}, \citenamefont {Johnson}, \citenamefont
  {Koretsune}, \citenamefont {Iba{\~{n}}ez-Azpiroz}, \citenamefont {Lee},
  \citenamefont {Lihm}, \citenamefont {Marchand}, \citenamefont {Marrazzo},
  \citenamefont {Mokrousov}, \citenamefont {Mustafa}, \citenamefont {Nohara},
  \citenamefont {Nomura}, \citenamefont {Paulatto}, \citenamefont
  {Ponc{\'{e}}}, \citenamefont {Ponweiser}, \citenamefont {Qiao}, \citenamefont
  {Thöle}, \citenamefont {Tsirkin}, \citenamefont {Wierzbowska}, \citenamefont
  {Marzari}, \citenamefont {Vanderbilt}, \citenamefont {Souza}, \citenamefont
  {Mostofi},\ and\ \citenamefont {Yates}}]{Pizzi20p165902}%
  \BibitemOpen
  \bibfield  {author} {\bibinfo {author} {\bibfnamefont {G.}~\bibnamefont
  {Pizzi}}, \bibinfo {author} {\bibfnamefont {V.}~\bibnamefont {Vitale}},
  \bibinfo {author} {\bibfnamefont {R.}~\bibnamefont {Arita}}, \bibinfo
  {author} {\bibfnamefont {S.}~\bibnamefont {Blügel}}, \bibinfo {author}
  {\bibfnamefont {F.}~\bibnamefont {Freimuth}}, \bibinfo {author}
  {\bibfnamefont {G.}~\bibnamefont {G{\'{e}}ranton}}, \bibinfo {author}
  {\bibfnamefont {M.}~\bibnamefont {Gibertini}}, \bibinfo {author}
  {\bibfnamefont {D.}~\bibnamefont {Gresch}}, \bibinfo {author} {\bibfnamefont
  {C.}~\bibnamefont {Johnson}}, \bibinfo {author} {\bibfnamefont
  {T.}~\bibnamefont {Koretsune}}, \bibinfo {author} {\bibfnamefont
  {J.}~\bibnamefont {Iba{\~{n}}ez-Azpiroz}}, \bibinfo {author} {\bibfnamefont
  {H.}~\bibnamefont {Lee}}, \bibinfo {author} {\bibfnamefont {J.-M.}\
  \bibnamefont {Lihm}}, \bibinfo {author} {\bibfnamefont {D.}~\bibnamefont
  {Marchand}}, \bibinfo {author} {\bibfnamefont {A.}~\bibnamefont {Marrazzo}},
  \bibinfo {author} {\bibfnamefont {Y.}~\bibnamefont {Mokrousov}}, \bibinfo
  {author} {\bibfnamefont {J.~I.}\ \bibnamefont {Mustafa}}, \bibinfo {author}
  {\bibfnamefont {Y.}~\bibnamefont {Nohara}}, \bibinfo {author} {\bibfnamefont
  {Y.}~\bibnamefont {Nomura}}, \bibinfo {author} {\bibfnamefont
  {L.}~\bibnamefont {Paulatto}}, \bibinfo {author} {\bibfnamefont
  {S.}~\bibnamefont {Ponc{\'{e}}}}, \bibinfo {author} {\bibfnamefont
  {T.}~\bibnamefont {Ponweiser}}, \bibinfo {author} {\bibfnamefont
  {J.}~\bibnamefont {Qiao}}, \bibinfo {author} {\bibfnamefont {F.}~\bibnamefont
  {Thöle}}, \bibinfo {author} {\bibfnamefont {S.~S.}\ \bibnamefont {Tsirkin}},
  \bibinfo {author} {\bibfnamefont {M.}~\bibnamefont {Wierzbowska}}, \bibinfo
  {author} {\bibfnamefont {N.}~\bibnamefont {Marzari}}, \bibinfo {author}
  {\bibfnamefont {D.}~\bibnamefont {Vanderbilt}}, \bibinfo {author}
  {\bibfnamefont {I.}~\bibnamefont {Souza}}, \bibinfo {author} {\bibfnamefont
  {A.~A.}\ \bibnamefont {Mostofi}}, \ and\ \bibinfo {author} {\bibfnamefont
  {J.~R.}\ \bibnamefont {Yates}},\ }\href {\doibase 10.1088/1361-648x/ab51ff}
  {\bibfield  {journal} {\bibinfo  {journal} {Journal of Physics: Condensed
  Matter}\ }\textbf {\bibinfo {volume} {32}},\ \bibinfo {pages} {165902}
  (\bibinfo {year} {2020})}\BibitemShut {NoStop}%
\bibitem [{\citenamefont {Lado}\ and\ \citenamefont
  {Fern{\'a}ndez-Rossier}(2016)}]{Lado16p035023}%
  \BibitemOpen
  \bibfield  {author} {\bibinfo {author} {\bibfnamefont {J.}~\bibnamefont
  {Lado}}\ and\ \bibinfo {author} {\bibfnamefont {J.}~\bibnamefont
  {Fern{\'a}ndez-Rossier}},\ }\href@noop {} {\bibfield  {journal} {\bibinfo
  {journal} {2D Materials}\ }\textbf {\bibinfo {volume} {3}},\ \bibinfo {pages}
  {035023} (\bibinfo {year} {2016})}\BibitemShut {NoStop}%
\bibitem [{\citenamefont {Rice}\ and\ \citenamefont
  {Mele}(1982)}]{Rice82p1455}%
  \BibitemOpen
  \bibfield  {author} {\bibinfo {author} {\bibfnamefont {M.}~\bibnamefont
  {Rice}}\ and\ \bibinfo {author} {\bibfnamefont {E.}~\bibnamefont {Mele}},\
  }\href@noop {} {\bibfield  {journal} {\bibinfo  {journal} {Physical Review
  Letters}\ }\textbf {\bibinfo {volume} {49}},\ \bibinfo {pages} {1455}
  (\bibinfo {year} {1982})}\BibitemShut {NoStop}%
\bibitem [{\citenamefont {Ishizuka}\ and\ \citenamefont
  {Nagaosa}(2017)}]{Ishizuka17p033015}%
  \BibitemOpen
  \bibfield  {author} {\bibinfo {author} {\bibfnamefont {H.}~\bibnamefont
  {Ishizuka}}\ and\ \bibinfo {author} {\bibfnamefont {N.}~\bibnamefont
  {Nagaosa}},\ }\href@noop {} {\bibfield  {journal} {\bibinfo  {journal} {New
  Journal of Physics}\ }\textbf {\bibinfo {volume} {19}},\ \bibinfo {pages}
  {033015} (\bibinfo {year} {2017})}\BibitemShut {NoStop}%
\bibitem [{\citenamefont {Kolpak}\ \emph {et~al.}(2006)\citenamefont {Kolpak},
  \citenamefont {Sai},\ and\ \citenamefont {Rappe}}]{Kolpak06p054112}%
  \BibitemOpen
  \bibfield  {author} {\bibinfo {author} {\bibfnamefont {A.~M.}\ \bibnamefont
  {Kolpak}}, \bibinfo {author} {\bibfnamefont {N.}~\bibnamefont {Sai}}, \ and\
  \bibinfo {author} {\bibfnamefont {A.~M.}\ \bibnamefont {Rappe}},\ }\href@noop
  {} {\bibfield  {journal} {\bibinfo  {journal} {Phys. Rev. B}\ }\textbf
  {\bibinfo {volume} {74}},\ \bibinfo {pages} {054112} (\bibinfo {year}
  {2006})}\BibitemShut {NoStop}%
\bibitem [{\citenamefont {Peierls}(1933)}]{Peierls33p763}%
  \BibitemOpen
  \bibfield  {author} {\bibinfo {author} {\bibfnamefont {R.}~\bibnamefont
  {Peierls}},\ }\href@noop {} {\bibfield  {journal} {\bibinfo  {journal}
  {Zeitschrift f{\"u}r Physik}\ }\textbf {\bibinfo {volume} {80}},\ \bibinfo
  {pages} {763} (\bibinfo {year} {1933})}\BibitemShut {NoStop}%
\bibitem [{\citenamefont {Fradkin}(2013)}]{Fradkin13field}%
  \BibitemOpen
  \bibfield  {author} {\bibinfo {author} {\bibfnamefont {E.}~\bibnamefont
  {Fradkin}},\ }\href@noop {} {\emph {\bibinfo {title} {Field theories of
  condensed matter physics}}}\ (\bibinfo  {publisher} {Cambridge University
  Press},\ \bibinfo {year} {2013})\BibitemShut {NoStop}%
\bibitem [{\citenamefont {Gao}\ \emph {et~al.}(2021)\citenamefont {Gao},
  \citenamefont {Addison}, \citenamefont {Mele},\ and\ \citenamefont
  {Rappe}}]{Gao21p042032}%
  \BibitemOpen
  \bibfield  {author} {\bibinfo {author} {\bibfnamefont {L.}~\bibnamefont
  {Gao}}, \bibinfo {author} {\bibfnamefont {Z.}~\bibnamefont {Addison}},
  \bibinfo {author} {\bibfnamefont {E.}~\bibnamefont {Mele}}, \ and\ \bibinfo
  {author} {\bibfnamefont {A.~M.}\ \bibnamefont {Rappe}},\ }\href@noop {}
  {\bibfield  {journal} {\bibinfo  {journal} {Physical Review Research}\
  }\textbf {\bibinfo {volume} {3}},\ \bibinfo {pages} {L042032} (\bibinfo
  {year} {2021})}\BibitemShut {NoStop}%
\bibitem [{\citenamefont {Zhang}\ \emph {et~al.}(2019)\citenamefont {Zhang},
  \citenamefont {Holder}, \citenamefont {Ishizuka}, \citenamefont {de~Juan},
  \citenamefont {Nagaosa}, \citenamefont {Felser},\ and\ \citenamefont
  {Yan}}]{Zhang19p3783}%
  \BibitemOpen
  \bibfield  {author} {\bibinfo {author} {\bibfnamefont {Y.}~\bibnamefont
  {Zhang}}, \bibinfo {author} {\bibfnamefont {T.}~\bibnamefont {Holder}},
  \bibinfo {author} {\bibfnamefont {H.}~\bibnamefont {Ishizuka}}, \bibinfo
  {author} {\bibfnamefont {F.}~\bibnamefont {de~Juan}}, \bibinfo {author}
  {\bibfnamefont {N.}~\bibnamefont {Nagaosa}}, \bibinfo {author} {\bibfnamefont
  {C.}~\bibnamefont {Felser}}, \ and\ \bibinfo {author} {\bibfnamefont
  {B.}~\bibnamefont {Yan}},\ }\href@noop {} {\bibfield  {journal} {\bibinfo
  {journal} {Nature Communications}\ }\textbf {\bibinfo {volume} {10}},\
  \bibinfo {pages} {3783} (\bibinfo {year} {2019})}\BibitemShut {NoStop}%
\bibitem [{\citenamefont {Kraut}\ and\ \citenamefont {von{
  }Baltz}(1979)}]{Kraut79p1548}%
  \BibitemOpen
  \bibfield  {author} {\bibinfo {author} {\bibfnamefont {W.}~\bibnamefont
  {Kraut}}\ and\ \bibinfo {author} {\bibfnamefont {R.}~\bibnamefont {von{
  }Baltz}},\ }\href@noop {} {\bibfield  {journal} {\bibinfo  {journal} {Phys.
  Rev. B}\ }\textbf {\bibinfo {volume} {19}},\ \bibinfo {pages} {1548}
  (\bibinfo {year} {1979})}\BibitemShut {NoStop}%
\bibitem [{\citenamefont {Morimoto}\ and\ \citenamefont
  {Nagaosa}(2016)}]{Morimoto16pe1501524}%
  \BibitemOpen
  \bibfield  {author} {\bibinfo {author} {\bibfnamefont {T.}~\bibnamefont
  {Morimoto}}\ and\ \bibinfo {author} {\bibfnamefont {N.}~\bibnamefont
  {Nagaosa}},\ }\href {\doibase 10.1126/sciadv.1501524} {\bibfield  {journal}
  {\bibinfo  {journal} {Science Advances}\ }\textbf {\bibinfo {volume} {2}},\
  \bibinfo {pages} {e1501524} (\bibinfo {year} {2016})}\BibitemShut {NoStop}%
\bibitem [{\citenamefont {Tan}\ and\ \citenamefont
  {Rappe}(2016)}]{Tan16p237402}%
  \BibitemOpen
  \bibfield  {author} {\bibinfo {author} {\bibfnamefont {L.~Z.}\ \bibnamefont
  {Tan}}\ and\ \bibinfo {author} {\bibfnamefont {A.~M.}\ \bibnamefont
  {Rappe}},\ }\href {\doibase 10.1103/PhysRevLett.116.237402} {\bibfield
  {journal} {\bibinfo  {journal} {Physical Review Letters}\ }\textbf {\bibinfo
  {volume} {116}},\ \bibinfo {pages} {237402} (\bibinfo {year}
  {2016})}\BibitemShut {NoStop}%
\end{thebibliography}%



\section*{Supplementary material}

# Magnetic Bulk Photovoltaic Effect: Strong and Weak Field

Zhenbang Dai[1] and Andrew M. Rappe[1]

[1]*Department of Chemistry, University of Pennsylvania,
Philadelphia, Pennsylvania 19104–6323, USA*
(Dated: June 24, 2022)

## HAMILTONIAN FOR THE MODEL SYSTEM

The Hamiltonian for such model is:

$$H = H^{RM} + H^L,$$
$$H^{RM} = \sum_{I,J}(-1)^I \Delta c_{I,J}^\dagger c_{I,J} - \sum_{I,J}\left(t_1 + (-1)^I dt_1\right) c_{I+1,J}^\dagger c_{I,J} - \sum_{I,J}(t_2 + (-1)^I dt_2)c_{I,J+1}^\dagger c_{I,J} + c.c., \quad (S1)$$
$$H^L = -\sum_{K,J}\delta d_{K,J}^\dagger d_{K,J} - \sum_{K,J} t_3 d_{K+1,J}^\dagger d_{K,J} - \sum_{K,J} t_3 d_{K,J+1}^\dagger d_{K,J} - \sum_{\{I,K\},J} t_3 d_{K,J}^\dagger c_{I,J} + c.c.. \quad (S2)$$

Eq. (S1) describes the central optically active region, where the first term is the onsite energy, the second term is the hopping along the $x$-direction, and the third term is the hopping along the $y$-direction. Eq. (S2) describes the leads, where each term has a similar meaning as in Eq. (S1), except the fourth term which describes the coupling between the leads and the central region. $I$ and $K$ are the site indices along the $x$-direction for central region and leads, respectively, and $J$ is the site index along $y$-direction. $\{I, K\}$ represents the sites at the interface.

## TEST OF OUR MODEL WITH LEADS

A test of our model is presented in Fig. (S1), which shows that the non-magnetic shift current from this model agrees with that calculated from a bulk system. Moreover, in the case where the magnetic field is commensurate with the size of the central region, i.e., the magnetic flux through the central region is equal to an integer number of flux quanta $h/e$, the Peierls phase will increase by $2n\pi$ from one end to another, restoring periodic boundary conditions for the magnetic unit cell. Thus, one can still calculate the current by adopting a periodic bulk system for specific magnetic field values. These results are in excellent agreement with the results from our model with leads for both longitudinal shift current and transverse mag-BC. Note that we only show the $xxX$ shift current and $xxY$ mag-BC because other components can be proven to vanish (See the following two sections). After showing the validity of our model, we use this model to explore in more detail how BPVE behaves at weak and strong magnetic field.

## DERIVATION OF THE MAGNETIC HAMILTONIAN IN BLOCH BASIS

In this section, we are going to derive the Hamiltonian when the magnetic field is commensurate with the central region, which will be used to calculate the shift current and mag-BC as shown in Fig. 1b and 1d. This will also be used to argue why the $xxY$ shift current and $xxX$ mag-BC will not exist when the magnetic field is not turned on.

Let's first assume the existence of a magnetic unit cell with lattice parameters $R_x^0$ and $NR_y^0$, and then derive the requirements that the magnetic field needs to fulfill for the existence of magnetic unit cell. It is known that the influence of the magnetic field can be included via the Peierls substitution, which will add an extra phase factor, the Peierls phase, to atomic orbital. So, the basis functions for the magnetic unit cell can be constructed as:

$$\Phi_I(\mathbf{x}) = \frac{1}{\sqrt{N}} \sum_{\mathbf{R}} e^{i\mathbf{k}\cdot(\mathbf{R}+\boldsymbol{\tau}_I)} \phi_I(\mathbf{x} - \boldsymbol{\tau}_I - \mathbf{R}), \quad (S3)$$

$$\phi_I(\mathbf{x} - \boldsymbol{\tau}_I - \mathbf{R}) = e^{i\int_{\mathbf{r}_I}^{\mathbf{x}} \mathbf{A_B}\cdot d\mathbf{r}} \phi_I^0(\mathbf{x} - \boldsymbol{\tau}_I - \mathbf{R}). \quad (S4)$$

Here, $\mathbf{R} = (mR_x^0, nNR_y^0)$ is a compact notation that runs over all the lattice vectors, $\boldsymbol{\tau}_I$ is the coordinate of I-th site in the first magnetic unit cell, $\mathbf{r}_I \equiv \mathbf{R} + \boldsymbol{\tau}_I$ and $\phi^0$ is the atomic orbital for non-magnetic system. All the other notations are consistent with the main text. To lighten the notation, we will first skip the $\mathbf{k}$-dependence in our derivation, and it will pose no difficulty when adding $\mathbf{k}$-dependence back. The Hamiltonian under this basis can be written as:

$$\langle \Phi_I | H | \Phi_I \rangle = \frac{1}{N}\sum_{\mathbf{R}}\sum_{\mathbf{R}'} e^{i\frac{e}{\hbar}B\frac{(r'_{I,y}-r_{I,y})(r_{I,x}+r'_{I,x})}{2}} \left\langle \phi_I^0(\mathbf{x}) |H_0| \phi_I^0(\mathbf{x} - \mathbf{r}'_I + \mathbf{r}_I)\right\rangle$$
$$= \frac{1}{N}\sum_{\mathbf{R}}\sum_{\mathbf{R}''} e^{i\frac{e}{\hbar}B\frac{R''_y R''_x}{2}} e^{i\frac{e}{\hbar}BR''_y r_{I,x}} \left\langle \phi_I^0(\mathbf{x})|H_0|\phi_I^0(\mathbf{x}-\mathbf{R}'')\right\rangle, \quad (S5)$$

2

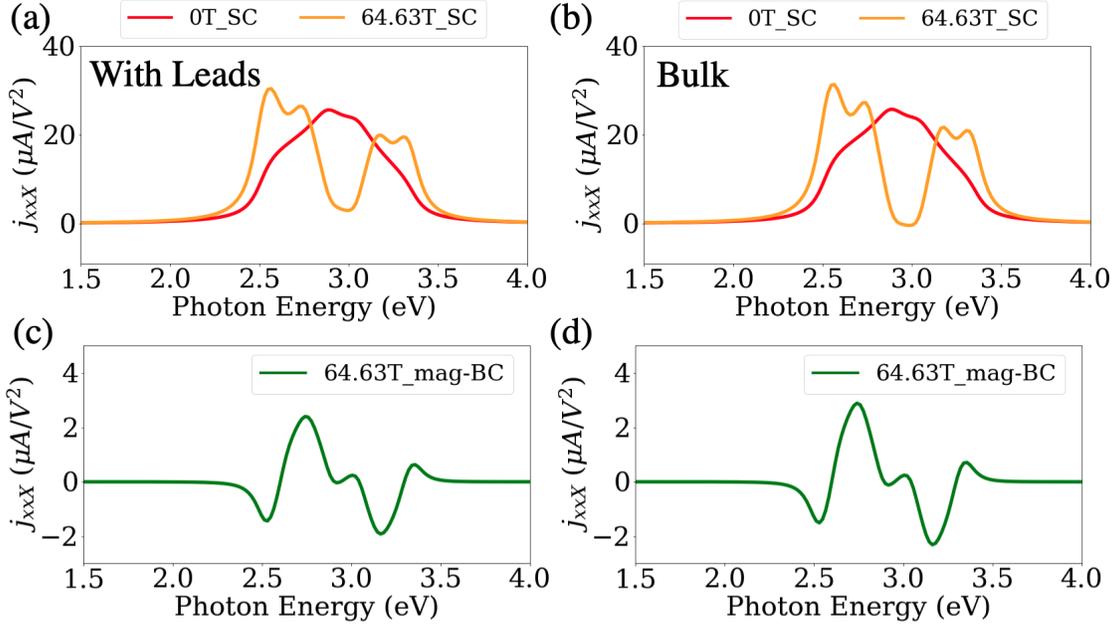

FIG. S1. Comparison of the various BPVE photocurrents from our model with electrodes and from the bulk system. (a) and (b): The $xxX$ component of the shift current without magnetic field and with magnetic field. (c) and (d): The $xxY$ component of the magnetically-induced ballistic current with magnetic field. The magnetic field is set to be commensurate with the 400 sites in the optically active region. The agreement between the leads model ((a) and (c)) and bulk system ((b) and (d)) is excellent, enabling us to investigate how BPVE reponse evolves as the magnetic field changes.

where the Peierls substitution (in Landau gauge) for a square lattice has been used:

$$\frac{e}{\hbar} \int_{\mathbf{R_i}}^{\mathbf{R_j}} \mathbf{A}_B \cdot d\mathbf{r} = \frac{e}{\hbar} B \frac{(R_{j,y} - R_{i,y})(R_{j,x} + R_{i,x})}{2}. \tag{S6}$$

Since $r_{I,x} = mR_x^0 + \tau_{I,x}$, it will have dependence over $\mathbf{R}$, which we want to eliminate so that the summation over $\mathbf{R}$ in Eq. S5 can be performed as in conventional tight-binding model with periodic boundary condition. Rewriting the term that is $\mathbf{R}$-dependent,

$$\exp\left(i\frac{e}{\hbar}BR_y'' r_{I,x}\right) = \exp\left(i\frac{e}{\hbar}Bmn''NR_x^0 R_y^0\right) \exp\left(i\frac{e}{\hbar}Bn''NR_y^0 \tau_{I,x}\right), \tag{S7}$$

we can find that in order to remove the $\mathbf{R}$-dependence, i.e., the m-dependence, we require $B$ to be a specific value such that:

$$\frac{e}{\hbar}Bmn''NR_x^0 R_y^0 = 2l\pi. \tag{S8}$$

In other words, to make the sure the phase factor Eq. S8 is always an integer multiple of $2\pi$, the minimal value for $B$ would be:

$$\frac{e}{\hbar}BNR_y^0 R_x^0 = 2\pi, \ B = \frac{1}{NR_x^0 R_y^0}\frac{2\pi\hbar}{e}. \tag{S9}$$

Substituting Eq. S9 into Eq.S5 and considering the $\mathbf{k}$-dependence, we can arrive at the following expression for the magnetic Hamiltonian:

$$\langle \Phi_{I,k}|H|\Phi_{I,k}\rangle = (-1)^I \Delta - 2\left(t_2 + (-1)^I dt_2\right)\cos\left(\frac{e}{\hbar}BR_y^0 \tau_{I,x} + k_y R_y^0\right), \tag{S10}$$

$$\langle \Phi_{I,k}|H|\Phi_{I+1,k}\rangle = \left(t_1 + (-1)^I dt_1\right) e^{ik_x \tau_{ab}}, \tag{S11}$$

where $\tau_{ab} = 4$ Å is the bond length. From the above two equations, we can see that for a given $k_y$, the magnetic band structure will be symmetric for $k_x$ and $-k_x$ because the Hamiltonians are conjugate of each other while the eigenvalues are always real for a Hermitian matrix. Thus, the contribution at $(k_x, k_y)$ and $(-k_x, k_y)$ to the $xxX$ component of Eq. (2) will cancel each other and thus give a zero mag-BC along longitudinal direction.

3

# EXPRESSIONS OF SHIFT CURRENT AND INJECTION CURRENT UNDER MAGNETIC FIELD

One may be concerned with the expression of shift current and injection current under magnetic field and wonder if they will change. So, in this section, we demonstrate that their expressions are invariant under the magnetic field.

The derivation of shift current and injection current requires an independent-particle unperturbed Hamiltonian and the electron-light interaction: $H = H_0 + H_{e-light}$. If we know the eigenstates and energies $\psi_{nk}$ and $\varepsilon_{nk}$, as well as the optical matrix $v_{nm}(k)$, then the shift current will take the familiar form as in [1]. In our case, the unperturbed Hamiltonian is the one with the the magnetic field: $H_0 = (p - eA_B)^2/2m + V(x)$, where $A_B$ is the vector potential of the static magnetic field, and the electron-light interaction has the same form of minimal coupling: $H_{e-light} = 1/m(p - eA_B) \cdot A_E$, where $A_E$ is the vector potential of the light. So, if we define the optical matrix as $v = 1/m(p - eA_B) = i/\hbar[H_0, x]$, then we can compute $v$ from tight-binding Hamiltonians of $H_0$ [2]. In the meantime, eigenstates and energies of $H_0$ can be obtained as well.

On the other hand, one may think that the breaking of time-reversal symmetry could influence the form the shift current exrepssion. But as proved by our previous work [3], the derivation of shift current does not rely on the existence of time-reversal symmetry, and for a semiconductor there is no so-called Fermi-surface contribution [3]. So, in the end, the functional form of shift current and injection current will not change, but all the ingredients will correspond to the magnetic Hamiltonian $H_0 = (p - eA_B)^2/2m + V(x)$.

# TRANSVERSE SHIFT CURRENT

In this section, we are going to show why the transverse (xxY) shift current will be zero even after turning on the magnetic field. The proof is based on magnetic unit cell, but the same argument can be made for our model with leads. For non-magnetic shift current where the time-reversal-symmetry (TRS) is preserved, the mirror symmetry about the $y = 0$ plane will make any transverse current vanish. However, when TRS is broken, this is not manifest, but the same conclusion can still be obtained by analyzing the structure of the shift current expression Eq. (1) and the magnetic Hamiltonian.

When calculating numerically, Eq. (1) is not the most convenient formula to use since the derivative of phase is not so easy to be evaluated directly, and a direct discretation of Eq. (1) will artificially break the gauge invariance. Instead, people usually use the logarithm derivative and start from the following quantity [1]:

$$-i\frac{\partial}{\partial k_q} \langle n\mathbf{k}|\hat{v}^r|l\mathbf{k}\rangle - [\chi_{lq}(\mathbf{k}) - \chi_{nq}(\mathbf{k})] \langle n\mathbf{k}|\hat{v}^r|l\mathbf{k}\rangle$$
$$= -\langle n\mathbf{k}|\hat{v}^r|l\mathbf{k}\rangle \lim_{\Delta k_q \to 0} i\frac{1}{\Delta k_q} \left[ \ln \frac{\langle n\mathbf{k}+\Delta k_q|\hat{v}^r|l\mathbf{k}+\Delta k_q\rangle}{\langle n\mathbf{k}|\hat{v}^r|l\mathbf{k}\rangle} + \ln \frac{\langle n\mathbf{k}|n\mathbf{k}+\Delta k_q\rangle}{\langle l\mathbf{k}|l\mathbf{k}+\Delta k_q\rangle} \right], \quad (S12)$$

and evaluate the following term as the shift vector instead of the second line of Eq. (5):

$$\lim_{\Delta k_q \to 0} -i\frac{1}{\Delta k_q} \left[ \ln \frac{\langle n\mathbf{k}+\Delta k_q|\hat{v}^r|l\mathbf{k}+\Delta k_q\rangle}{\langle n\mathbf{k}|\hat{v}^r|l\mathbf{k}\rangle} + \ln \frac{\langle n\mathbf{k}|n\mathbf{k}+\Delta k_q\rangle}{\langle l\mathbf{k}|l\mathbf{k}+\Delta k_q\rangle} \right]. \quad (S13)$$

For the transverse (xxY) shift current, $q = y$ and $r = x$. To evaluate the velocity matrix $\langle n\mathbf{k}|\hat{v}^x|l\mathbf{k}\rangle$, we need to calculate the matrix $\frac{\partial}{\partial k_x} \langle \Phi_{I,\mathbf{k}}|H_0|\Phi_{I',\mathbf{k}}\rangle$. For magnetic unit cell where the lattice periodicity along $x$-direction is very large, it is sufficient to do the analysis for a $\mathbf{k}$-grid where $k_x$ is always equal to zero because the first Brillouin zone is very narrow along the $b_x$-direction. In doing so, the diagonal elements of $\frac{\partial}{\partial k_x} \langle \Phi_{I,\mathbf{k}}|H_0|\Phi_{I',\mathbf{k}}\rangle$ will always be zero because Eq. (S10) has no dependence on $k_x$, while the off-diagonal elements will be imaginary as can be told from Eq. (S11). Thus, $\frac{\partial}{\partial k_x} \langle \Phi_{I,\mathbf{k}}|H_0|\Phi_{I',\mathbf{k}}\rangle$ will be a purely imaginary matrix. In addition, the magnetic Hamiltonian as defined in Eq. (S10) and Eq. (S11) is always real and symmetric, which means that the eigenvectors will also be real no matter what $k_y$ is.

Now we look at the first terms in Eq. S13. Since $|n\mathbf{k}\rangle$ and $|l\mathbf{k}\rangle$ are real and the $\frac{\partial}{\partial k_x} \langle \Phi_{I,\mathbf{k}}|H_0|\Phi_{I',\mathbf{k}}\rangle$ is imaginary, it can be seen from [2] that both $\langle n\mathbf{k}+\Delta k_q|\hat{v}^r|l\mathbf{k}+\Delta k_q\rangle$ and $\langle n\mathbf{k}|\hat{v}^r|l\mathbf{k}\rangle$ will be imaginary, but $\frac{\langle n\mathbf{k}+\Delta k_q|\hat{v}^r|l\mathbf{k}+\Delta k_q\rangle}{\langle n\mathbf{k}|\hat{v}^r|l\mathbf{k}\rangle}$ will be real. For the second term in Eq. S13, since $|n\mathbf{k}\rangle$ and $|n\mathbf{k}+\Delta k_y\rangle$ are real vectors, $\frac{\langle n\mathbf{k}|n\mathbf{k}+\Delta k_q\rangle}{\langle l\mathbf{k}|l\mathbf{k}+\Delta k_q\rangle}$ is also a real quantity. As a result, Eq. S13 will be imaginary, but according to Eq. (1), the shift vector needs to be real in order to get a non-vanishing shift current (at least for light polarization along $x$-direction). Therefore, the transverse shift current (xxY) will always be zero even when the magnetic field is turned on, no matter whether the field is the strong or weak.

## ABSORPTION COEFFICIENT

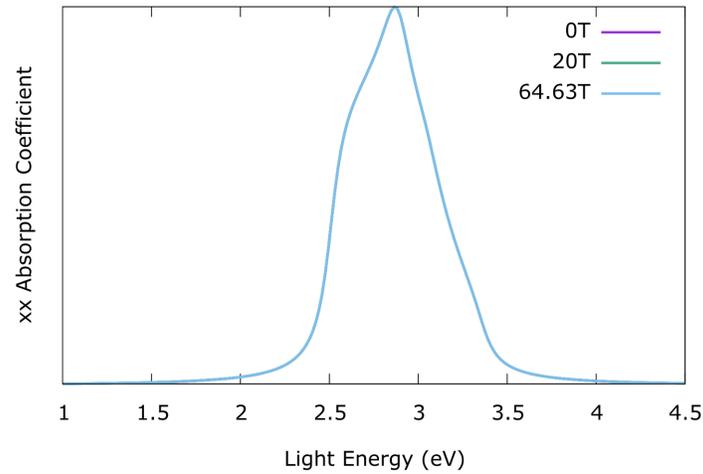

FIG. S2. $xx$ component of the absoprtion coefficient.

Unlike the $xxX$ shift current which exhibits a drastic change from weak field to strong field, the optical absorption barely changes as can be seen in Fig. . This is because the absorption coefficient does not depend on the phase of wave functions, while the shift vector makes the shift current depend sensitively on the phase. Thus, the induced Perierls phase will manifest in shift current but not in the absoprtion coefficient.

## BENCHMARK WITH MORE ATOMS IN LEADS

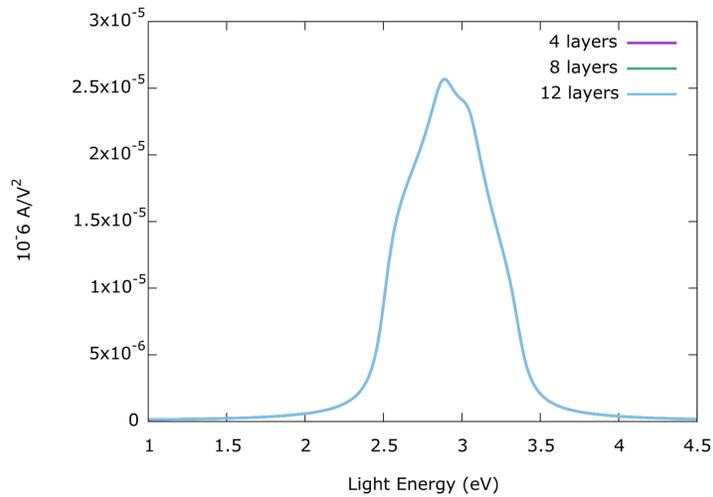

FIG. S3. Shift current for different layers of atoms in leads without magnetic field. Adding more sites in the leads will not change the final results.

# BENCHMARK WITH THE MAGNETIC FIELD EXTENDED TO THE LEADS

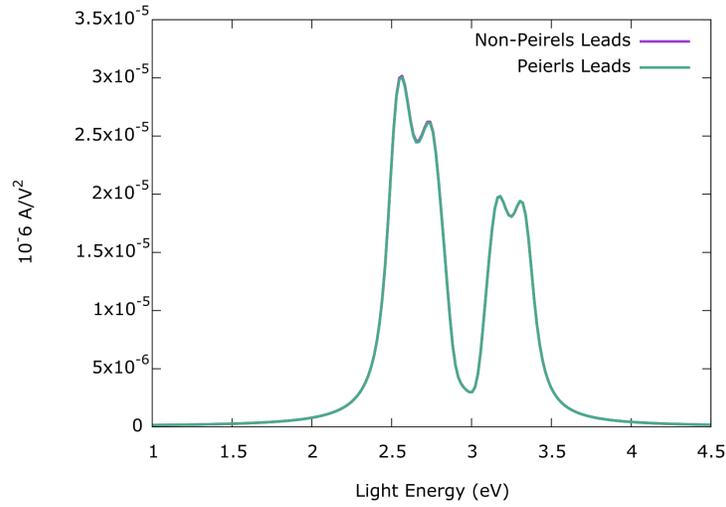

FIG. S4. Shift current for different ways of treating the magnetic field. The magnetic field is commensurate with 400 Rice-Mele sites, which is consistent with the choice made in the main text. Whether or not extending the magnetic field into the leads will create negligible impact, as can be seen from the almost perfect match between the Non-Peierls Leads (magnetic field in the leads) and Peierls Leads (no magnetic field in the leads). Despite the negligible influence, it should be noted that the two treatments correspond to two different physical scenarios. It is because we are mainly concerned with the response from the central region that makes the two treatments give the same results.

# EVOLUTION OF MAG-BC FOR A HIGHER RESOLUTION OF MAGNETIC FIELD

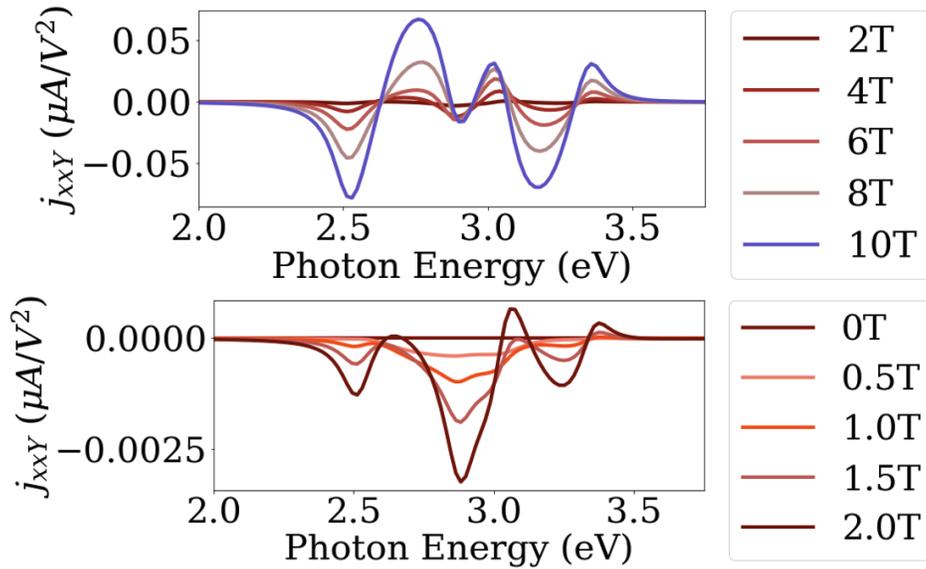

FIG. S5. Evolution of the mag-BC from 0 T to 10 T.

# EVOLUTION OF BAND STRUCTURES

6

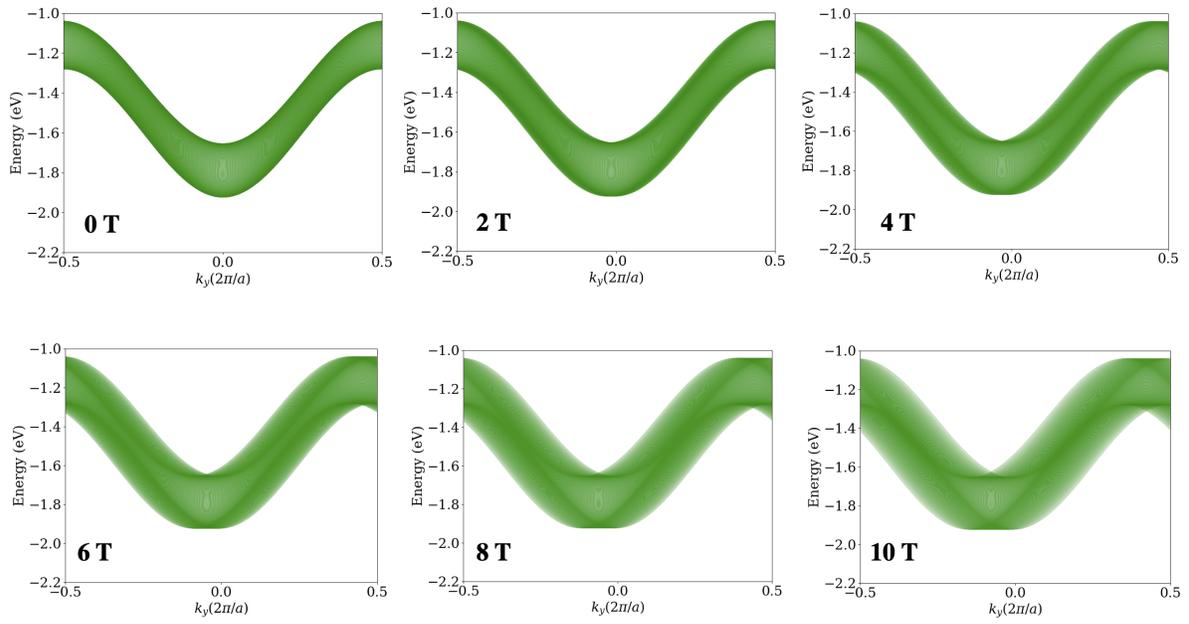

FIG. S6. Evolution of the valence bands from 0 T to 10 T.

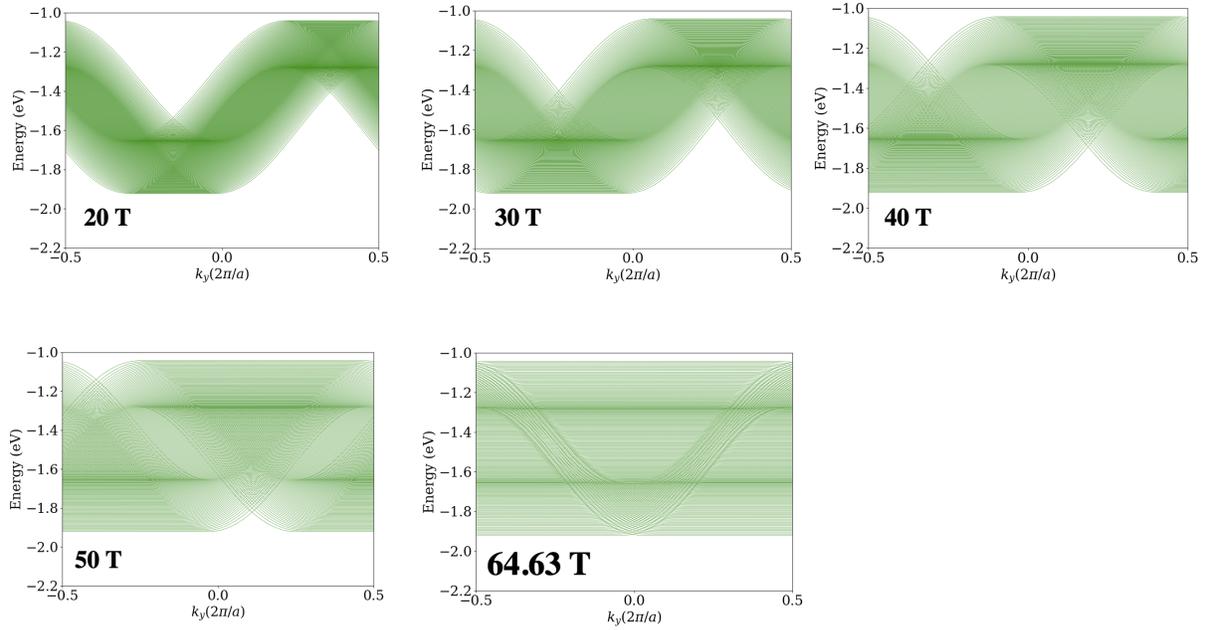

FIG. S7. Evolution of the valence bands from 20 T to 64.63 T.

[1] S. M. Young and A. M. Rappe, Phys. Rev. Lett. **109**, 116601 (2012).
[2] T. G. Pedersen, K. Pedersen, and T. B. Kriestensen, Physical Review B **63**, 201101 (2001).
[3] L. Gao, Z. Addison, E. Mele, and A. M. Rappe, Physical Review Research **3**, L042032 (2021).


\end{document}